\documentstyle[11pt,newpasp,twoside,epsf]{article} 
\def\edcomment#1{\iffalse\marginpar{\raggedright\sl#1\/}\else\relax\fi} 
\reversemarginpar 

\begin{document} 
\title{Multiwavelength Time Series Data of the LMC Cluster Reticulum}
\author{Monelli$^1$ M., Andreuzzi$^2$ G., Bono$^2$ G., Buonanno$^1$ R.,
Caputo$^2$ F., Castellani$^2$ M., Castellani$^2$ V., Corsi$^2$ C. E.,
Dall' Ora$^1$ M., Marconi G.$^3$, Pulone$^2$ L., Ripepi$^4$ V., Storm$^5$ J.,
Testa$^2$ V.}
\affil{{}$^1$ Universit\`a di Roma Tor Vergata}
\affil{{}$^2$ INAF-Osservatorio Astronomico di Roma} 
\affil{{}$^3$ ESO-Chile}
\affil{{}$^4$ INAF-Osservatorio Astronomico di Napoli}
\affil{{}$^5$ Postdam Observatory}
\begin{abstract} 
We present accurate multiwavelength UBVI time series data of the LMC 
cluster Reticulum. Data cover a time interval of $\approx6$ yr and
have been collected with SUSI1/2 and SOFI at NTT/ESO. 
For each band we collected approximately 30 short/long exposures
and the total exposure times range from roughly 3500 (U, B), 6000s (V) to
8300s (I). The observing strategy and data reduction (DAOPHOTII/ALLFRAME)
allowed us to reach a photometry accuracy of 0.02 magnitude from the tip
of the Red Giant Branch well below the Turn-Off region.
Even though this cluster presents a very low central density
($\rho_0 = 1.0 M_{\sun}/pc^3$), we found a sizable sample of Blue
Stragglers (BSs). 
We also selected stars with a variability index larger than 2, and 
interestingly enough we detected together with the RR Lyrae stars a 
large sample of variable stars around and below the TO region. 
Preliminary analysis on the luminosity variation indicate that 
these objects might be binary stars.
\end{abstract}
\section{Preliminary results} 
The LMC cluster Reticulum presents several interesting 
features, such as low-reddening (E(B-V) = 0.03 $\pm$ 0.02), low-central 
density ($\rho_0 = 1.0 M_{\sun}/pc^3$), accurate 
metallicity estimates ([Fe/H] = -1.71 $\pm$ 0.09), 
and a sizable sample of RR Lyrae stars (Walker, 1992). 
Moreover and even more importantly it is marginally affected by
field contamination since it is located at approximately 11$^o$ from
the LMC bar (l =269$^o$, b=-40$^o$). 
In spite of these undisputable advantages only a few 
investigations have been devoted to this cluster. In fact, after the 
pioneering work by Gratton \& Ortolani (1987) accurate photometry for 
both the RR Lyrae and the bright static stars was provided by Walker (1992).
In this paper we present some preliminary results concerning the Reticulum 
stellar content well below the Turn-Off (TO) region. Data plotted in
the figure show quite clearly that this cluster shares the same
evolutionary features of old LMC clusters (Olsen et al., 1998). 
In this context it is worth mentioning the sizable sample of BSs we
detected. They 
can be easily identified in the V, B-V and in the U, U-V diagram. Thanks to 
the accuracy of current data it seems that they can be split into a hotter 
and a cooler group. The former one is located at $V\approx21.5$ and 
$B-V\approx0.25$ while the latter at $V\approx21.5$ and 
$B-V\approx0.5$ mag. Finally we note that BSs in the U, U-B diagram 
appear, as expected, cooler than sub giant branch stars
\footnote{This work has been supported by the MURST/Cofin2000 under
the project: Stellar observables of cosmological relevance.}.
\begin{figure}
\plotfiddle{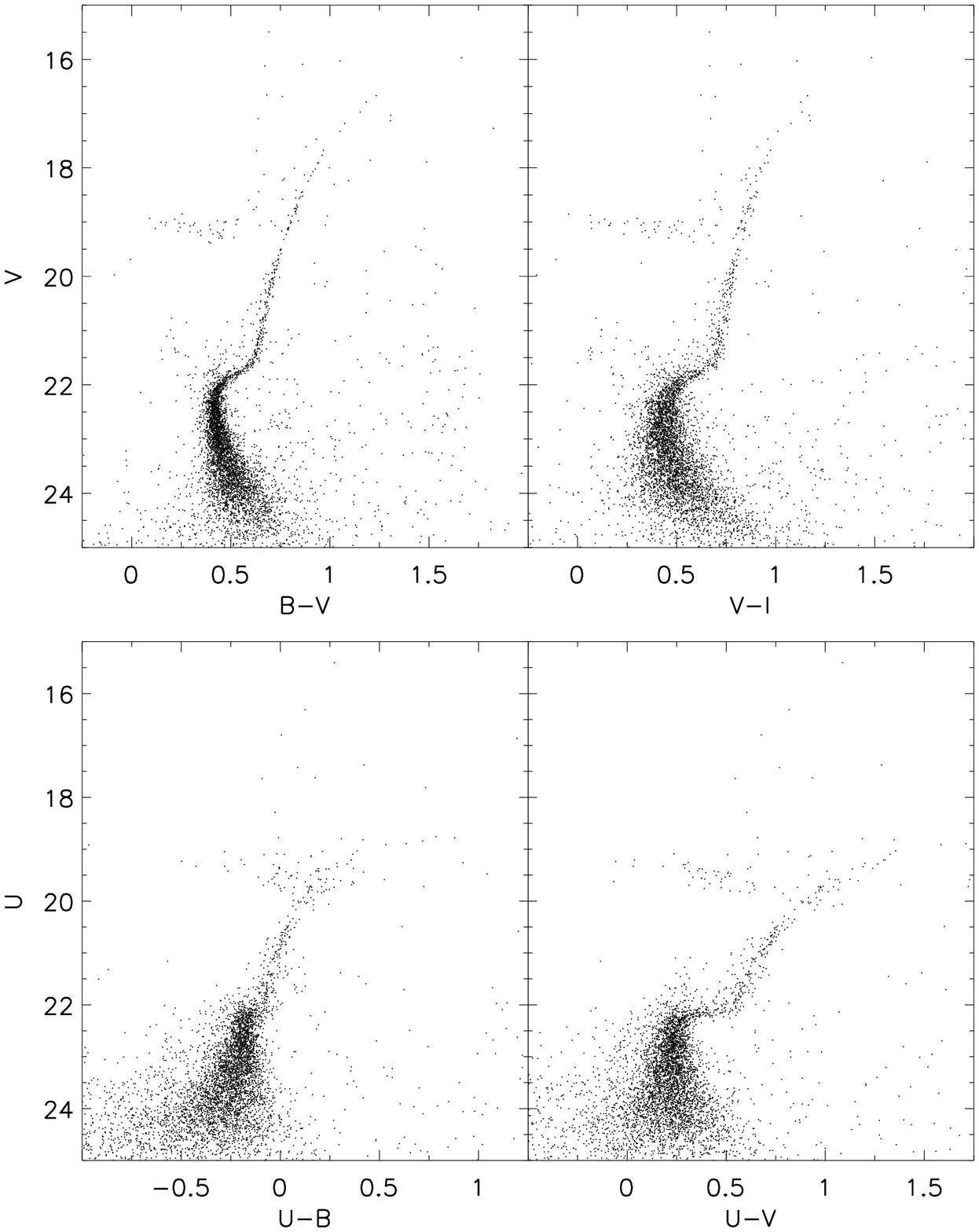}{12.cm}{0}{58}{58}{-190/in}{-85/in}
\end{figure} 


\begin{references}
Gratton, R. G., Ortolani, S., 1987, \aaps, 71, 131\\
Olsen, K. A. G., Hodge, P. W., Mateo, M., Olszewski, E. W., Schommer,
R. A., Suntzeff, N. B., Walker, A. R., 1998, \mnras, 300, 665\\
Walker, A. R., 1992, \aj, 103, 1166
\end{references}
\end{document}